\documentclass[reprint,aps,pra,amssymb,amsmath,amsfonts,onecolumn]{revtex4-2}

\usepackage[utf8]{inputenc}
\usepackage[T1]{fontenc}
\usepackage{microtype}

\usepackage{graphicx}
\usepackage{xcolor}
\usepackage[colorlinks=true,linkcolor=blue,urlcolor=blue,citecolor=blue,bookmarks=true,bookmarksopenlevel=2]{hyperref}

\usepackage{array,tabularx,booktabs}
\usepackage{rotating}
\usepackage{mhchem}
\usepackage{xspace}
\usepackage{siunitx}
\newcommand{\cf}[1]{\ensuremath{\mathrm{#1}}}
\usepackage{soul}

\begin{document}

\title{\textsc{Rci-q}: an improved QED correction model for the \textsc{Grasp2018} package}
\author{Karol Kozio{\l}}
\email{Karol.Koziol@ncbj.gov.pl}
\affiliation{Narodowe Centrum Bada\'{n} J\k{a}drowych (NCBJ), Andrzeja So{\l}tana 7, 05-400 Otwock-\'{S}wierk, Poland}

\begin{abstract}
The \textsc{Rci-Q} package is an extension to the \textsc{Grasp2018} suite, improving the model of estimating the quantum-electrodynamics corrections to the energy levels. 
The Flambaum--Ginges radiative potential method is used to estimate the leading self-energy correction to electron energy in many electron atoms. The new fitting prefactors to parameterize radiative potential are presented. The correction to self-energy originating from finite nucleus size is included. The Wichmann-Kroll part of the vacuum polarization potential is also implemented. 
\end{abstract}

\maketitle



\section{Introduction}
\label{sec:introduction}

Quantum electrodynamics (QED) corrections to the energy levels are important for higher-$Z$ highly charged atoms \cite{Indelicato2019}. 
While the leading-order vacuum polarization correction (correction to the electron--nucleus interaction photon propagator, related to the creation and annihilation of virtual electron-positron pairs in the field of the nucleus) is easily implemented in atomic calculations as a Uehling potential \cite{Uehling1935}, using the self energy correction (correction to the electron propagator, arising from the interaction of the electron with its own radiation field) in atomic calculations is much more complicated. Leading self-energy radiative corrections in hydrogen-like systems were evaluated firstly by Bethe \cite{Bethe1947} within the non-relativistic approach, then relativistically by Mohr \cite{Mohr1974}. There are a couple of methods to estimate the self energy corrections to energy levels in multi-electron atomic systems. 
The default model of estimating the self-energy correction implemented in the \texttt{rci} program of the \textsc{Grasp2018} package \cite{FroeseFischer2019} is simple, but not accurate in some cases \cite{Koziol2018a}. The alternative methods that can be working combined with the \textsc{Grasp2018} programs, such as the \texttt{rci4} program \cite{Lowe2013,Nguyen2023}, based on the Welton concept, and the \textsc{Qedmod} package \cite{Shabaev2015}, based on model Lamb-shift operator, provide better numbers, but using them requires additional steps. So, there is a need for self-energy correction implementation in \textsc{Grasp2018}, that is calculated on-the-fly during regular \texttt{rci} run. 
The Flambaum--Ginges radiative potential method \cite{Flambaum2005,Ginges2016} was implemented recently in the \textsc{amb}i\textsc{t} program \cite{Kahl2019a}, the \textsc{Dirac25} suite \cite{Saue2025}, and the modified versions of the \textsc{Grasp} \cite{Thierfelder2010,Piibeleht2022} and \textsc{Turbomole} \cite{Janke2025} packages. 
In the present work the Flambaum--Ginges radiative potential method is incorporated into a modified version of the \texttt{rci} program, called \texttt{rci-q}. The updated fitting prefactors to parameterize radiative potential with a wide range of atomic number $Z$ and quantum numbers $n$ and $l$ for subshells are presented. These prefactors may be also used in the other implementations of Flambaum--Ginges potential in atomic and molecular codes. 
Moreover, the finite nucleus size correction to self-energy is added. The Wichmann-Kroll part of the vacuum polarization potential is also implemented. 
Because the QED effects are more important for high-$Z$ atoms, it is expected that the \texttt{rci-q} program will be more suitable for calculations involving heavy atoms, than the original \texttt{rci} program. 

\section{Theoretical background}
\label{sec:theory}

The Flambaum--Ginges radiative potential method \cite{Flambaum2005,Ginges2016} is used to estimate the leading self-energy correction to electron energy in many electron atoms. The self-energy radiative potential is expressed as 
\begin{equation}
V_\text{SE}(r) = V_\text{el}(r)+V_\text{mag}(r)+V_\text{low}(r)
\end{equation}
where $V_\text{el}(r)$ is the high-frequency part of electric contribution (called later as electric part), $V_\text{low}(r)$ is the low-frequency part of electric contribution (called later as low-frequency part), and $V_\text{mag}(r)$ is the magnetic part. 

The derivation of such potential has been made in Refs.~\cite{Flambaum2005,Berestetskii1982} for the first order interaction between electron and virtual photon in the electric field of the nucleus. The radiative potential related to this interaction can be divided into three parts (marked by curly parentheses in the equation below):
\begin{equation}
\begin{array}{>{\displaystyle}l>{\displaystyle }l}
V_\text{rad}(r) = & \left\{ \frac{2\alpha}{3\pi} V_\text{nuc}(r) \int_{1}^{\infty} \frac{\sqrt{t^2-1}}{t^2} \left( 1+\frac{1}{2t^2} \right) e^{-2tr/\alpha} \;\text{d}t \right\} \\[3ex]
& + \left\{ - \frac{\alpha}{\pi} V_\text{nuc}(r) \int_{1}^{\infty} \frac{e^{-2tr/\alpha}}{\sqrt{t^2-1}} \left[ \left( 1-\frac{1}{2t^2} \right) \times \left( \ln(t^2-1) + \ln(4\alpha^2/\lambda^2) \right) +\frac{1}{t^2}-\frac{3}{2} \right] \text{d}t \right\} \\[3ex]
& + \left\{ \frac{i\alpha^2}{4\pi} \vec\gamma\cdot\vec\nabla \left[ V_\text{nuc}(r) \int_{1}^{\infty} \frac{e^{-2tr/\alpha}}{t^2\sqrt{t^2-1}} \;\text{d}t -1 \right] \right\}
\end{array}
\end{equation}
The first term is the well-known Uehling potential, leading vacuum polarization term, dependent on nuclear potential. The second term is a part of self-energy dependent on nuclear potential and the low-frequency cutoff parameter $\lambda$. It is known as the electric part of self-energy. Finally, the third term is a part of self-energy dependent on nuclear potential divergence. At long distances it describes the interaction of the electron anomalous magnetic moment with the atomic electrostatic potential \cite{Berestetskii1982} -- so, it is known as the magnetic part of the self-energy. 
For the sake of computational efficiency and removing the cutoff parameter from the equation, the electric part can be divided into the high-frequency part, which dominates for $l=0$ orbitals, and the low-frequency part, which dominates for $l>0$ orbitals.

\subsection{Point-like nucleus}

For point-like nucleus ($V_\text{nuc}(r) = Z/r$) the electric part is expressed as 
\begin{equation}
\begin{array}{>{\displaystyle}l>{\displaystyle }l}
V_\text{el}(r) = & A(Z,n,l,r)\; \frac{\alpha}{\pi} V_\text{nuc}(r) 
\int_{1}^{\infty} \frac{e^{-2tr/\alpha}}{\sqrt{t^2-1}} \left[ \left( 1-\frac{1}{2t^2} \right) \right.\\[3ex] & \left. \times \left( \ln(t^2-1) + 4\ln\left(\frac{1}{Z\alpha}+\frac{1}{2}\right) \right) +\frac{1}{t^2}-\frac{3}{2} \right] \text{d}t
\end{array}
\end{equation}
parametrized by a coefficient $A(Z,n,l,r)$. 
This equation may be also written in more convenient form as 
\begin{equation}
\begin{array}{>{\displaystyle}l>{\displaystyle }l}
V_\text{el}(r) = & A(Z,n,l,r)\; \frac{\alpha}{\pi} \frac{Z}{r} 
\left\{ 4\ln\left(\frac{1}{Z\alpha}+\frac{1}{2}\right) \left[Ki_0\left(\frac{2r}{\alpha}\right)-\frac{1}{2}Ki_2\left(\frac{2r}{\alpha}\right)\right] \right. 
\\[3ex]
& \left. + \left[Ki_0^\text{ln}\left(\frac{2r}{\alpha}\right)-\frac{1}{2}Ki_2^\text{ln}\left(\frac{2r}{\alpha}\right)+Ki_2\left(\frac{2r}{\alpha}\right)-\frac{3}{2}Ki_0\left(\frac{2r}{\alpha}\right)\right] \right\}
\end{array}
\end{equation}
Then the magnetic part for point-like nucleus is expressed as  
\begin{equation}
\begin{array}{>{\displaystyle}l>{\displaystyle }l}
V_\text{mag}(r) & = \frac{i\alpha^2}{4\pi} \vec\gamma\cdot\vec\nabla \left[ V_\text{nuc}(r) \int_{1}^{\infty} \frac{e^{-2tr/\alpha}}{t^2\sqrt{t^2-1}} \;\text{d}t -1 \right]
\\[3ex]
& = \frac{i\alpha^2}{4\pi} \vec\gamma\cdot\frac{\vec r}{r} \frac{Z}{r^2} \left[ \int_{1}^{\infty} \frac{e^{-2tr/\alpha}}{t^2\sqrt{t^2-1}} \;\text{d}t + \frac{2r}{\alpha}\int_{1}^{\infty} \frac{e^{-2tr/\alpha}}{t\sqrt{t^2-1}} \;\text{d}t -1 \right]
\end{array}
\end{equation}
This equation may be written in more convenient form as 
\begin{equation}
\begin{array}{>{\displaystyle}l>{\displaystyle }l}
V_\text{mag}(r) = i\vec\gamma\cdot\frac{\vec r}{r} \tilde{V}_\text{mag}(r)\\[3ex]
\tilde{V}_\text{mag}(r) = \frac{\alpha^2}{4\pi} \frac{Z}{r^2} \left[ Ki_2\left(\frac{2r}{\alpha}\right) + \frac{2r}{\alpha}Ki_1\left(\frac{2r}{\alpha}\right) -1 \right]
\end{array}
\end{equation}

\begin{table}[!t]
\caption{Fitting coefficients for the $A(Z,n,l)$ prefactor for $l=0$ orbitals. The fitting function is $f(Z) = a_0+a_1Z+a_2Z^2+a_3Z^3+a_4Z^4$.\label{tab:AZ}}
\begin{tabular*}{\linewidth}{@{} @{\extracolsep{\fill}} l *5{S[table-format=2.5e2]} @{}}
\toprule
$n$ & {$a_0$} & {$a_1$} & {$a_2$} & {$a_3$} & {$a_4$} \\
\midrule
\multicolumn{6}{c}{$Z\ge20$}\\
1 & 7.72308e-01 & -2.40991e-04 & 3.48842e-05 & -2.83516e-07 & 3.16093e-10 \\
2 & 8.07899e-01 & 1.29047e-03 & 3.94430e-05 & -1.64904e-07 & -1.97988e-09 \\
3 & 8.08505e-01 & 2.07848e-03 & 3.83122e-05 & -1.64870e-07 & -2.26659e-09 \\
4 & 8.08957e-01 & 2.52039e-03 & 3.03563e-05 & -5.24535e-08 & -2.92995e-09 \\
$\ge5$ & 8.15199e-01 & 2.19164e-03 & 4.33051e-05 & -2.04857e-07 & -2.39770e-09 \\
\midrule
\multicolumn{6}{c}{$Z<20$}\\
1 & 8.72587e-01 & -1.44109e-02 & -1.48436e-03 & 3.37161e-04 & -1.34952e-05 \\
2 & 8.91352e-01 & -1.26732e-02 & -1.29823e-03 & 3.24461e-04 & -1.32155e-05 \\
3 & 8.81396e-01 & 3.14301e-02 & -1.49689e-02 & 1.71096e-03 & -5.74309e-05 \\
4 & 8.83234e-01 & 3.17102e-02 & -1.49614e-02 & 1.71022e-03 & -5.74210e-05 \\
$\ge5$ & 8.84127e-01 & 3.18795e-02 & -1.49649e-02 & 1.71050e-03 & -5.74349e-05 \\
\bottomrule
\end{tabular*}
\end{table}

The integrals 
\begin{equation}Ki_n(x) = \int_{1}^{\infty} \frac{e^{-xt}}{t^n\sqrt{t^2-1}} \;\text{d}t\end{equation} 
are the Bickley–Naylor functions (note that $Ki_0(x)$ is also the modified Bessel function of the second kind and the zeroth order). 
The integrals 
\begin{equation}Ki_n^\text{ln}(x) = \int_{1}^{\infty} \frac{e^{-xt}\ln(t^2-1)}{t^n\sqrt{t^2-1}} \;\text{d}t\end{equation} 
are similar to $Ki_n(x)$ but contain additional logarithmic part. 

The $Ki_n(x)$ and $Ki_n^\text{ln}(x)$ integrals are calculated by using closed Newton--Cotes formula of fourth order (Boole's rule) on the exponential grid $t(i) = a e^{i b}$, $i = 1, \ldots, i_\text{max}$, with parameters $a = \num{1.0e-10}$, $b = 0.15$, and $i_\text{max}=300$. 
Although the number of grid points is small, the mean relative error of calculating integrals is \num{3.8e-3} comparing to the integrals calculated on a grid having 100-times more points ($a = \num{1.0e-10}$, $b = 0.0015$, and $i_\text{max}=30000$). The slightly lower accuracy is an acceptable price for the better code performance. Integrals calculated on the denser grid were compared to the ones determined by Janke et al. (supplementary material in Ref.~\cite{Janke2025}), obtaining \num{7.9e-4} mean relative difference.

Finally, the low frequency part is expressed as
\begin{equation}
V_\text{low}(r) = B(Z,n,\kappa)\; Z^4 \alpha^3 e^{-Zr}
\end{equation}
parametrized by a coefficient $B(Z,n,\kappa)$. 

\begin{table}[!t]
\caption{Fitting coefficients for the $B(Z,n,\kappa)$ prefactor for $l=1$ orbitals. The fitting function is $f(Z) = a_0+a_1(Z\alpha)+a_2(Z\alpha)^2+a_3(Z\alpha)^3$.\label{tab:BZp}}
\begin{tabular*}{\linewidth}{@{} @{\extracolsep{\fill}} l *4{S[table-format=2.5e2]} @{}}
\toprule
$n$ & {$a_0$} & {$a_1$} & {$a_2$} & {$a_3$} \\
\midrule
\multicolumn{5}{c}{$n\text{p}_{1/2}$, $\kappa=1$, $Z\ge25$}\\
2 & 3.00078e-02 & 3.10841e-01 & -1.05901e-01 & 1.40965e-01 \\
3 & 3.56110e-03 & 6.47205e-01 & -6.50979e-01 & 4.75313e-01 \\
4 & 3.25045e-02 & 5.33089e-01 & -4.02912e-01 & 3.16655e-01 \\
$\ge5$ & 3.92048e-02 & 5.33916e-01 & -3.96387e-01 & 3.03409e-01 \\
\midrule
\multicolumn{5}{c}{$n\text{p}_{1/2}$, $\kappa=1$, $Z<25$}\\
2 & -1.94268e-01 & 4.69211e+00 & -2.78403e+01 & 5.74180e+01 \\
3 & -2.11216e-01 & 6.45208e+00 & -4.61844e+01 & 1.11580e+02 \\
4 & -2.12246e-01 & 7.02072e+00 & -5.26197e+01 & 1.32414e+02 \\
$\ge5$ & -2.11575e-01 & 7.27284e+00 & -5.55193e+01 & 1.41980e+02 \\
\midrule
\multicolumn{5}{c}{$n\text{p}_{3/2}$, $\kappa=-2$, $Z\ge25$}\\
2 & 6.19859e-02 & 1.46532e-02 & 6.18175e-01 & -4.26690e-01 \\
3 & 7.37296e-02 & 1.61238e-01 & 3.87553e-01 & -2.56977e-01 \\
4 & 1.00974e-01 & 3.20842e-02 & 7.16133e-01 & -4.70388e-01 \\
$\ge5$ & 1.04827e-01 & 4.38918e-02 & 7.07673e-01 & -4.61889e-01 \\
\midrule
\multicolumn{5}{c}{$n\text{p}_{3/2}$, $\kappa=-2$, $Z<25$}\\
2 & 1.85706e-01 & -1.16736e+00 & 2.29089e+00 & 6.11143e+00 \\
3 & 2.40357e-01 & -1.93270e+00 & 6.61815e+00 & 1.23795e+00 \\
4 & 2.61959e-01 & -2.28189e+00 & 9.18272e+00 & -3.77179e+00 \\
$\ge5$ & 2.72890e-01 & -2.45901e+00 & 1.05770e+01 & -6.80477e+00 \\
\bottomrule
\end{tabular*}
\end{table}

\begin{table}[!t]
\caption{Fitting coefficients for the $B(Z,n,\kappa)$ prefactor for $l=2,3$ orbitals. The fitting function is $f(Z) = a_0+a_1(Z\alpha)+a_2(Z\alpha)^2+a_3(Z\alpha)^3+a_4(Z\alpha)^4$.\label{tab:BZdf}}
\begin{tabular*}{\linewidth}{@{} @{\extracolsep{\fill}} l *5{S[table-format=2.5e2]} @{}}
\toprule
$n$ & {$a_0$} & {$a_1$} & {$a_2$} & {$a_3$} & {$a_4$} \\
\midrule
\multicolumn{6}{c}{$n\text{d}_{3/2}$, $\kappa=2$, $Z\ge30$}\\
3 & -3.81685e-01 & 3.29496e+00 & -8.71608e+00 & 1.06372e+01 & -4.67039e+00 \\
4 & 1.35414e-01 & -8.79351e-01 & 2.89313e+00 & -2.96666e+00 & 1.10994e+00 \\
$\ge5$ & 2.12121e-01 & -1.65418e+00 & 5.79399e+00 & -7.41881e+00 & 3.48481e+00 \\
\midrule
\multicolumn{6}{c}{$n\text{d}_{3/2}$, $\kappa=2$, $Z<30$}\\
3 & -2.64617e-01 & -2.52618e+00 & 4.91292e+01 & -2.05113e+02 & 2.76993e+02 \\
4 & -1.95345e-01 & -3.21072e+00 & 6.37902e+01 & -2.98788e+02 & 4.45729e+02 \\
$\ge5$ & -1.70173e-01 & -3.53693e+00 & 7.10608e+01 & -3.48097e+02 & 5.42911e+02 \\
\midrule
\multicolumn{6}{c}{$n\text{d}_{5/2}$, $\kappa=-3$, $Z\ge30$}\\
3 & 6.10299e-01 & -3.49392e+00 & 8.56384e+00 & -8.66208e+00 & 3.26329e+00 \\
4 & 5.21836e-01 & -3.64388e+00 & 1.11067e+01 & -1.37627e+01 & 6.17809e+00 \\
$\ge5$ & 3.28721e-01 & -2.09216e+00 & 6.74317e+00 & -8.63612e+00 & 4.04192e+00 \\
\midrule
\multicolumn{6}{c}{$n\text{d}_{5/2}$, $\kappa=-3$, $Z<30$}\\
3 & 2.80273e-01 & 3.81077e-01 & -3.24881e+00 & -1.91187e+01 & 7.07767e+01 \\
4 & 2.46212e-01 & 5.19477e-01 & -4.94680e+00 & -1.98439e+01 & 9.31935e+01 \\
$\ge5$ & 2.36708e-01 & 6.17387e-01 & -6.28049e+00 & -1.68757e+01 & 9.79283e+01 \\
\midrule
\multicolumn{6}{c}{$n\text{f}_{5/2}$, $\kappa=3$, $Z\ge30$}\\
4 & -5.12300e+00 & 2.27980e+01 & -3.13854e+01 & 1.46058e+01 & \\
$\ge5$ & -3.59484e+00 & 1.94922e+01 & -3.33546e+01 & 1.90586e+01 & \\
\midrule
\multicolumn{6}{c}{$n\text{f}_{5/2}$, $\kappa=3$, $Z<30$}\\
4 & -2.06959e+00 & -8.36718e-01 & 5.99394e+00 & 5.48666e+01 & \\
$\ge5$ & -1.33925e+00 & -1.45686e+00 & 1.22530e+01 & 3.43523e+01 & \\
\midrule
\multicolumn{6}{c}{$n\text{f}_{7/2}$, $\kappa=-4$, $Z\ge30$}\\
4 & 3.05544e+00 & -5.39201e+00 & 5.93283e-02 & 3.70609e+00 & \\
$\ge5$ & 3.23661e+00 & -1.22567e+01 & 1.75365e+01 & -7.87039e+00 & \\
\midrule
\multicolumn{6}{c}{$n\text{f}_{7/2}$, $\kappa=-4$, $Z<30$}\\
4 & 1.97971e+00 & -5.34157e-01 & 9.03007e+00 & -4.34423e+01 & \\
$\ge5$ & 1.38843e+00 & -5.30541e-01 & 8.95192e+00 & -4.53794e+01 & \\
\bottomrule
\end{tabular*}
\end{table}

For one-electronic Dirac bispinor
\begin{equation}
\psi_{n \kappa j m_j}(r,\theta,\phi) = 
\frac{1}{r}
\begin{pmatrix}
P_{n\kappa}(r)\cdot \Omega_{\kappa j}^{m_j}(\theta,\phi)\\[1ex]
i Q_{n\kappa}(r)\cdot \Omega_{-\kappa j}^{m_j}(\theta,\phi)
\end{pmatrix}
\end{equation}
the energy shift arising from the electric parts are \cite{Ginges2016}
\begin{equation}
\delta E_{n\kappa}^\text{SE,el} = \int_0^\infty \left[P_{n\kappa}^2(r)+Q_{n\kappa}^2(r)\right] V_\text{el}(r)\;\text{d}r
\end{equation}
and
\begin{equation}
\delta E_{n\kappa}^\text{SE,low} = \int_0^\infty \left[P_{n\kappa}^2(r)+Q_{n\kappa}^2(r)\right] V_\text{low}(r)\;\text{d}r
\end{equation}
while the energy shift arising from the magnetic part is
\begin{equation}
\delta E_{n\kappa}^\text{SE,mag} = 2 \int_0^\infty \left[P_{n\kappa}(r)Q_{n\kappa}(r)\right] \tilde{V}_\text{mag}(r)\;\text{d}r
\end{equation}
Total energy shift arising from self-energy is then
\begin{equation}
\delta E_{n\kappa}^\text{SE} = \delta E_{n\kappa}^\text{SE,el} + \delta E_{n\kappa}^\text{SE,mag} + \delta E_{n\kappa}^\text{SE,low}
\end{equation}
Conveniently, the self-energy correction to energy levels is presented in the form 
\begin{equation}
\delta E_{n\kappa}^\text{SE} = \frac{Z^4 \alpha^3}{\pi n^3} \, F(Z\alpha)
\end{equation}
where $F(Z\alpha)$ is a function depending slowly on $Z$.

\begin{table}[!t]
\caption{Fitting coefficients for the FNS correction to the self-energy. The fitting function is $f(Z) = a_1 e^{a_2Z\alpha} + a_3 e^{a_4(Z\alpha)^2}$. $FNS(Z,n,\kappa)$ is given in terms of $F(Z\alpha)$. \label{tab:FNS}}
\begin{tabular*}{\linewidth}{@{} @{\extracolsep{\fill}} l *4{S[table-format=2.5e2]} @{}}
\toprule
$n$ & {$a_1$} & {$a_2$} & {$a_3$} & {$a_4$} \\
\midrule
\multicolumn{5}{c}{$n\text{s}_{1/2}$, $\kappa=-1$, $10\le Z \le30$}\\
1 & -3.09277e-05 & 9.64416e+00 & 3.24443e-05 & -1.43637e+01 \\
2 & -2.79601e-05 & 1.01704e+01 & 2.89087e-05 & -1.33097e+01 \\
3 & -2.74527e-05 & 1.01718e+01 & 2.82109e-05 & -1.33317e+01 \\
4 & -2.72748e-05 & 1.01410e+01 & 2.79295e-05 & -1.34101e+01 \\
5 & -2.70461e-05 & 1.01414e+01 & 2.73706e-05 & -1.31494e+01 \\
\midrule
\multicolumn{5}{c}{$n\text{s}_{1/2}$, $\kappa=-1$, $Z>30$}\\
1 & -2.76256e-05 & 9.32479e+00 & -1.91156e-05 & 1.16065e+01 \\
2 & -1.86926e-05 & 1.04577e+01 & -3.21015e-05 & 1.20893e+01 \\
3 & -1.83170e-05 & 1.04458e+01 & -3.58678e-05 & 1.18642e+01 \\
4 & -1.85589e-05 & 1.03697e+01 & -3.72856e-05 & 1.16841e+01 \\
5 & -1.87158e-05 & 1.03075e+01 & -3.85170e-05 & 1.15438e+01 \\
\midrule
\multicolumn{5}{c}{$n\text{p}_{1/2}$, $\kappa=1$, $60\le Z \le90$}\\
2 & 5.26335e-05 & -7.17750e-01 & -6.74612e-06 & 1.30961e+01 \\
3 & 5.91651e-05 & -4.96987e-01 & -8.36193e-06 & 1.30431e+01 \\
4 & 6.08334e-05 & -4.02436e-01 & -9.10911e-06 & 1.29279e+01 \\
5 & 6.17536e-05 & -3.65755e-01 & -9.53345e-06 & 1.28341e+01 \\
\midrule
\multicolumn{5}{c}{$n\text{p}_{1/2}$, $\kappa=1$, $Z>90$}\\
2 & 4.42640e-11 & 2.50460e+01 & -5.87890e-06 & 1.40698e+01 \\
3 & 5.50159e-11 & 2.47977e+01 & -7.80443e-06 & 1.37860e+01 \\
4 & 6.16946e-11 & 2.45698e+01 & -8.76597e-06 & 1.35649e+01 \\
5 & 6.50837e-11 & 2.44369e+01 & -9.27414e-06 & 1.34268e+01 \\
\midrule
\multicolumn{5}{c}{$n\text{p}_{3/2}$, $\kappa=-2$, $Z>75$}\\
2 & -2.41782e-07 & 1.07560e+01 & 5.66075e-09 & 1.59279e+01 \\
3 & -2.12046e-07 & 1.20219e+01 & 2.54147e-06 & 9.82270e+00 \\
4 & -1.99391e-07 & 1.22111e+01 & 1.99792e-06 & 1.02062e+01 \\
5 & -1.93886e-07 & 1.22805e+01 & 1.62241e-06 & 1.04782e+01 \\
\midrule
\multicolumn{5}{c}{$n\text{d}_{3/2}$, $\kappa=2$, $Z>100$}\\
3 & -1.16135e-10 & 1.71090e+01 &  &  \\
4 & -1.84156e-10 & 1.71129e+01 &  &  \\
5 & -2.27131e-10 & 1.70925e+01 &  &  \\
\bottomrule
\end{tabular*}
\end{table}

\subsection{Fitting $A(Z)$ and $B(Z)$ coefficients}

The coefficients $A(Z,n,l,r) = A(Z,n,l)\left[r/(r+0.07Z^2\alpha^3)\right]$ and $B(Z,n,\kappa)$ are the fitting factors and they were found by fitting the self-energy shift for hydrogenlike ions. Originally, the $A(Z)$ and $B(Z)$ coefficients were set to reproduce the self-energy shift for $n=5$ subshells \cite{Flambaum2005,Ginges2016}, so they are not fit well for inner shells. The $A(Z)$ factors for $l=0$ were improved substantially by Thierfelder and Schwerdtfeger \cite{Thierfelder2010} and then they fit well for $s$-subshells for $Z>20$.  

In the present work the $A(Z,n,l)$ coefficients for $l=0$ were fitted to reference self-energy shift for hydrogenlike ions by fourth order polynomials $A(Z) = a_0+a_1Z+a_2Z^2+a_3Z^3+a_4Z^4$, separately for $n=1,2,3,4,5$. For $n>5$ it is assumed $A(Z,n>5,l=0)=A(Z,n=5,l=0)$, which is a reasonable assumption as the $A(Z,n,l=0)$ factors saturate with $n$ increases \cite{Thierfelder2010}. The fitting was performed separately for arbitrary selected $Z\ge20$ and $Z<20$ ranges, to decrease the number of fitting parameters. The $B(Z)$ coefficients for $l=0$ were set to $B(Z,l=0) = 0.074 + 0.35Z\alpha$, as originally suggested in the work of Flambaum and Ginges \cite{Flambaum2005}, because in the case of $l=0$ the electric part of self-energy outranks the low-energy part. 
The reference values for the self-energy shift for hydrogenlike ions used in the present work are: Mohr \cite{Mohr1992a} for 1s, 2s, 2p$_{1/2}$, and 2p$_{3/2}$ subshells in the range $5\le Z \le110$; Yerokhin et al. \cite{Yerokhin2025} for (3--5)s, (3--5)p$_{1/2}$, (3--5)p$_{3/2}$, (3--5)d$_{3/2}$, (3--5)d$_{5/2}$, (4--5)f$_{5/2}$, and (4--5)f$_{7/2}$ subshells in the range $5\le Z \le100$; Mohr and Kim \cite{Mohr1992} for (3--5)s, (3--5)p$_{1/2}$, (3--5)p$_{3/2}$, (3--5)d$_{3/2}$ subshells for $Z=110$; Le Bigot et al. \cite{LeBigot2001} for (3--5)d$_{5/2}$, (4--5)f$_{5/2}$, and (4--5)f$_{7/2}$ subshells for $Z=110$. 
Then, the $A(Z)$ values to be fitted are calculated as follow. Firstly, the trial \texttt{rci-q} runs with $A(Z)$ suppressed (i.e., $A(Z)=1$) and default Flambaum and Ginges $B(Z)$ values (marked $B(Z)=\text{FG}$) have been performed. For these calculations the self-energy shift is given by
\begin{equation}\label{eq:azfit1}
\delta E^\text{SE}_{A(Z)=1,B(Z)=\text{FG}} = \delta E^\text{SE,el}_{A(Z)=1} + \delta E^\text{SE,mag} + \delta E^\text{SE,low}_{B(Z)=\text{FG}}
\end{equation}
For proper $A(Z)$ values, the self-energy shift, calculated by using the Flambaum--Ginges potential, should be equal to the reference one (taken explicitly from the literature tables or the cubic interpolation of literature data): 
\begin{equation}\label{eq:azfit2}
\delta E^\text{SE}_{\text{ref}} = A(Z) \cdot \delta E^\text{SE,el}_{A(Z)=1} + \delta E^\text{SE,mag} + \delta E^\text{SE,low}_{B(Z)=\text{FG}}
\end{equation}\label{eq:azfit3}
Then, the $A(Z)$ coefficient can be obtained from the equation
\begin{equation}
A(Z) = \frac{\delta E^\text{SE}_\text{ref} - \delta E^\text{SE}_{A(Z)=1,B(Z)=\text{FG}}}{\delta E^\text{SE,el}_{A(Z)=1}}+1
\end{equation}
and further fitted by the polynomial. 
The fitting coefficients for the $A(Z)$ prefactors are stored in the Table~\ref{tab:AZ}. 
For $l=1$ the $A(Z)$ coefficient is used in the form $A(Z,n,l=1) = 1.071 - 1.976x^2 - 2.128x^3 + 0.169x^4$ (where $x=(Z-80)\alpha$) and for $l>1$ the $A(Z,n,l>1)$ coefficient is set to zero. These values are taken from the original work of Flambaum and Ginges \cite{Flambaum2005}. 

The $B(Z,n,\kappa)$ coefficients for $l>0$ were fitted by the third order polynomials $B(Z) = a_0+a_1(Z\alpha)+a_2(Z\alpha)^2+a_3(Z\alpha)^3$ for $l=1,3$ and by the fourth order polynomials $B(Z) = a_0+a_1(Z\alpha)+a_2(Z\alpha)^2+a_3(Z\alpha)^3+a_4(Z\alpha)^4$ for $l=2$ subshells, separately for arbitrary selected low-$Z$ and high-$Z$ ranges. 
Similar like in the case of $A(Z)$, the $B(Z)$ coefficients are calculated by using the trial \texttt{rci-q} runs with $B(Z)$ suppressed (i.e., $B(Z)=1$) and default Flambaum and Ginges $A(Z)$ values (marked $A(Z)=\text{FG}$). Then
\begin{equation}\label{eq:bzfit1}
\delta E^\text{SE}_{A(Z)=\text{FG},B(Z)=1} = \delta E^\text{SE,el}_{A(Z)=\text{FG}} + \delta E^\text{SE,mag} + \delta E^\text{SE,low}_{B(Z)=1}
\end{equation}
and 
\begin{equation}\label{eq:bzfit2}
\delta E^\text{SE}_{\text{ref}} = \delta E^\text{SE,el}_{A(Z)=\text{FG}} + \delta E^\text{SE,mag} + B(Z) \cdot \delta E^\text{SE,low}_{B(Z)=1}
\end{equation}
and the $B(Z)$ coefficient can be obtained from the equation
\begin{equation}\label{eq:bzfit3}
B(Z) = \frac{\delta E^\text{SE}_\text{ref} - \delta E^\text{SE}_{A(Z)=\text{FG},B(Z)=1}}{\delta E^\text{SE,low}_{B(Z)=1}}+1
\end{equation}
For $n>5$ it is assumed $B(Z,n>5,\kappa)=B(Z,n=5,\kappa)$, assuming the $B(Z,n,\kappa)$ factors saturate with $n$ increases. 
For subshells $l>3$ the $B(Z,n,\kappa)$ prefactor is set to zero. 
The fitting coefficients for the $B(Z)$ prefactors are stored in the Tables~\ref{tab:BZp} and~\ref{tab:BZdf}.

\begin{figure}[!htb]
\includegraphics[width=0.5\linewidth]{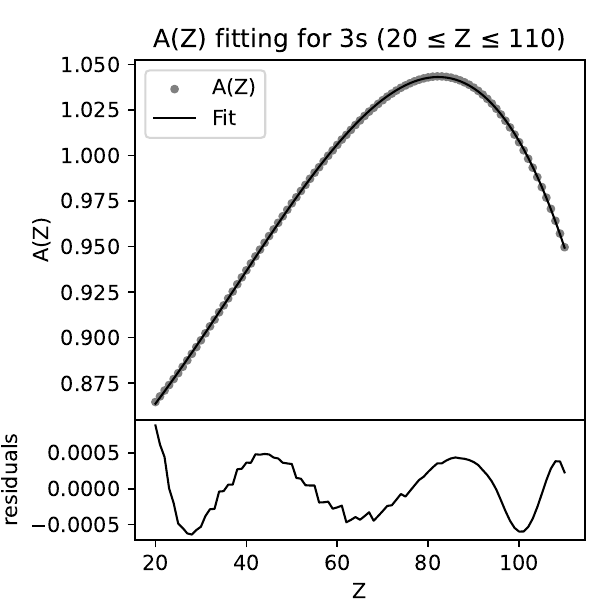}\hfill\includegraphics[width=0.5\linewidth]{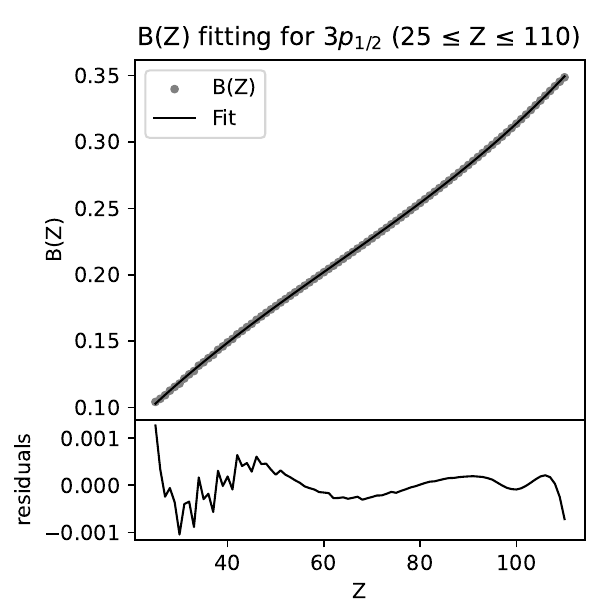}\\
\includegraphics[width=0.5\linewidth]{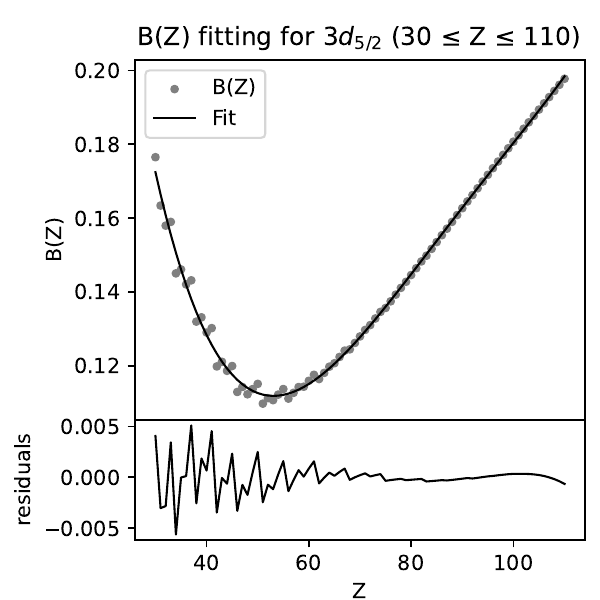}\hfill\includegraphics[width=0.5\linewidth]{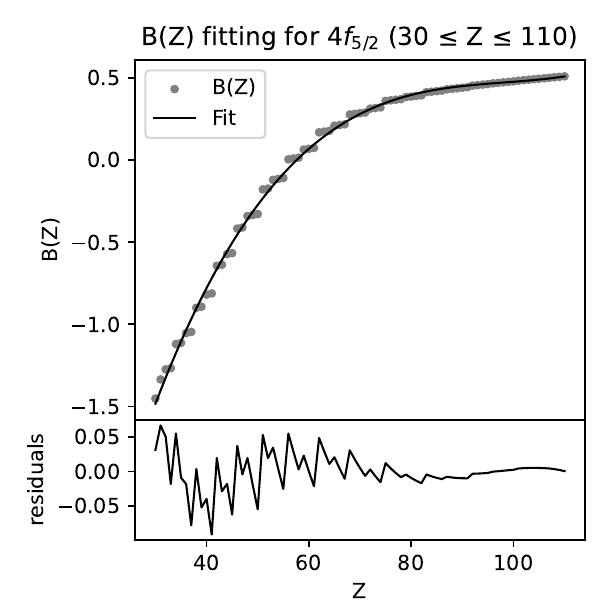}
\caption{Fitting plots for $A(Z)$ and $B(Z)$ coefficients for selected subshells.}\label{fig:residuals}
\end{figure}

The fitting plots for $A(Z)$ and $B(Z)$ coefficients, containing the residuals, for selected subshells are presented on Fig~\ref{fig:residuals}. Because $A(Z) = 0$ for $l\ge2$ subshells, the $B(Z)$ coefficients calculated by Eq.~\eqref{eq:bzfit3} are having a less smooth, steplike, $Z$-dependence, what makes the fit quality get worse.

\subsection{Finite nucleus size correction}
\begin{figure}[!htb]
\centering
\includegraphics[width=0.5\linewidth]{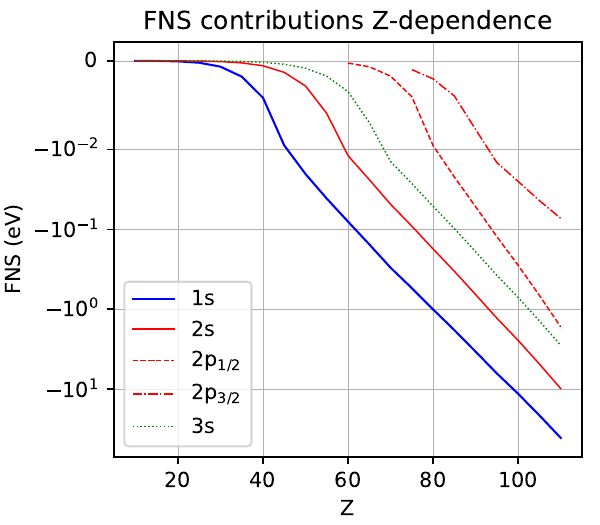}
\caption{$Z$-dependence of finite nucleus size correction to self-energy contribution to orbital energy for selected subshells.}\label{fig:fns}
\end{figure}

The correction to self-energy originating from finite nucleus size (FNS) is given by fitting the data taken from Refs.~\cite{Yerokhin2011,Shabaev2013}. The fitting was performed by function $FNS(Z,n,\kappa) = a_1 e^{a_2Z\alpha} + a_3 e^{a_4(Z\alpha)^2}$ for subshells (1--5)s separately for ranges $10\le Z \le30$ and $30< Z \le120$, for subshells (2--5)p$_{1/2}$ separately for ranges $60\le Z \le90$ and $90< Z \le120$, for subshells (2--5)p$_{3/2}$ in the range $75\le Z \le120$ and by function $FNS(Z,n,\kappa) = a_1 e^{a_2Z\alpha}$  for subshells (3--5)d$_{3/2}$ in the range $100\le Z \le120$. $FNS(Z,n,\kappa)$ is given in terms of $F(Z\alpha)$ in these fitting functions. The fitting coefficients for the FNS correction are stored in the Table~\ref{tab:FNS}. 
Below the $Z$ fitting range (i.e., $Z<10$ for (1--5)s, $Z<60$ for (2--5)p$_{1/2}$, $Z<75$ for (2--5)p$_{3/2}$, $Z<100$ for (3--5)d$_{3/2}$), the FNS correction is set to zero. For $Z>120$ range the FNS corrections are extrapolated by using coefficients fitted for ranges $Z\le120$. The $Z$-dependence of FNS contribution for selected subshells is presented on Fig.~\ref{fig:fns}.

\subsection{Wichmann-Kroll vacuum polarization correction}

The Wichmann-Kroll part of the vacuum polarization potential (terms of order higher than $\alpha(Z\alpha)$, i.e. $\alpha^2(Z\alpha)$, $\alpha(Z\alpha)^3$, etc.) has been implemented into the \textsc{rci-q} code according to Fainshtein et al. algorithm \cite{Fainshtein1991}. 
In that approach the Wichmann-Kroll vacuum polarization potential for the point-like nucleus is given as (in atomic units):
\begin{equation}
V^\text{(3+)}(r,Z) = 
\begin{cases}
	\displaystyle
V^\text{(3)}(r) \left( 1 + \sum_{n=1}^{4}\sum_{m=0}^{7} a_{0mn}\lambda_{n}r^{m+S_{0n}} \right)^{-1} \qquad \text{for\ \ } r \le 0.15 \\[1.5em]
	\displaystyle
\left( V^\text{(3)}(r) + \frac{Q^\text{(5+)}}{r} \right) \left( 1 + \sum_{n=1}^{3}\sum_{m=0}^{6} a_{1mn}\lambda_{n}r^{m+S_{1n}} \right)^{-1} \qquad \text{for\ \ } 0.15\le r \le 1.5 \\[1.5em]
	\displaystyle
V^\text{(3)}(r) \left( 1 + \sum_{n=1}^{3}\sum_{m=0}^{6} a_{2mn}\lambda_{n}r^{m+S_{2n}} \right)^{-1} \qquad \text{for\ \ } 1.5\le r \le 4 \\[1.5em]
	\displaystyle
V^\text{(3)}(r) \left( 1 + \sum_{n=1}^{2}\sum_{m=0}^{6} a_{3mn}\lambda_{n}r^{m+S_{3n}} \right)^{-1} \qquad \text{for\ \ } 4\le r \le 10 \\[1.5em]
	\displaystyle
\frac{1}{\pi\alpha} \sum_{m=0}^{4} \frac{(m!)^2}{r^{2m+5}} \sum_{n=1}^{[m/2]+1} f_{mn}(\alpha Z)^{2n+1} \qquad \text{for\ \ }r\ge10
\end{cases}
\end{equation}
where $[x]$ is an integral part of $x$ and $\lambda = 1-\sqrt{1-(\alpha Z)^2}$. 
The $Q^\text{(5+)}$ factor is given as:
\begin{equation}
Q^\text{(5+)} = \alpha^2 Z^3 \sum_{n=1}^{15} q_{n} \lambda^{n}
\end{equation}
and the $V^\text{(3)}(r)$ is given as:
\begin{equation}
V^\text{(3)}(r) =
\begin{cases}
	\displaystyle
\alpha^2 Z^3 \left( \sum_{n=0}^{8} p_{n}r^{n-1} + \sum_{n=0}^{5} p_{n+9}r^{n+2} \ln{r} + \sum_{n=0}^{4} p_{n+15}r^{n+3} \ln^2{r} \right) \quad \text{for\ \ } r \le 4.25 \\[1.5em]
	\displaystyle
\alpha^2 Z^3 \frac{1}{r^5} \sum_{n=0}^{15} p_{n+20} \left(\frac{10}{r}\right)^{2n} \quad \text{for\ \ } r > 4.25
\end{cases}
\end{equation}
The coefficients $a_{imn}$ and $S_{in}$ ($i=0,1,2,3$), $f_{mn}$, $q_{n}$, and $p_{n}$ are tabulated in the paper of Fainshtein et al. \cite{Fainshtein1991}. 

Because in the original \texttt{rci} program, the K{\"a}ll{\'e}n--Sabry part of vacuum polarization potential of order $\alpha^2(Z\alpha)$ is implemented, that part of code is commented in the \texttt{rci-q} program. The implemented Wichmann-Kroll vacuum polarization potential also covers the K{\"a}ll{\'e}n--Sabry contribution.

Comparison of present calculated Wichmann-Kroll vacuum polarization contribution to the energy of 1s, 2s, 2p$_{1/2}$, and 2p$_{3/2}$ subshells with literature data is presented in Table~\ref{tab:wk_comp}. It can be seen from that table that the present calculated values systematically overestimate the literature ones (Refs.~\cite{Persson1993}, \cite{Mohr1998}, and \cite{Ivanov2025}). For 1s and 2s subshells, when the Wichmann-Kroll contribution is the most important, the overestimation is in the range of 3--10\%. That systematic overestimation has been attributed to the finite nucleus size effect on the Wichmann-Kroll contribution \cite{Fainshtein1991}, because in the cited comparative references the uniform charge distribution for the nucleus was used, in difference to the Fainshtein et al. algorithm.

\begin{table}[!htb]
\caption{Wichmann-Kroll vacuum polarization contribution to the energy of 1s, 2s, 2p$_{1/2}$, and 2p$_{3/2}$ subshells calculated for selected hydrogenlike ions (labeled TW), in terms of eV, compared to literature data.}\label{tab:wk_comp}
\centering
\tabcolsep=0.5\tabcolsep
\begin{tabular*}{\linewidth}{@{}@{\extracolsep{\fill}} r *8{S[table-format=1.4]} @{}}
\toprule
& \multicolumn{4}{c}{1s} & \multicolumn{4}{c}{2s} \\
\cmidrule{2-5}\cmidrule{6-9}
$Z$ & {TW} & {Ref. \cite{Persson1993}} & {Ref. \cite{Mohr1998}} & {Ref. \cite{Ivanov2025}} & {TW} & {Ref. \cite{Persson1993}} & {Ref. \cite{Mohr1998}} & {Ref. \cite{Ivanov2025}} \\
\midrule
36 & 0.0160 & 0.0155 & 0.02 &  & 0.00205 & 0.00200 & 0.0020 &  \\
54 & 0.1740 & 0.1695 & 0.17 &  & 0.0235 & 0.0230 & 0.023 &  \\
70 & 0.8567 & 0.8283 & 0.824 & 0.8318 & 0.1240 & 0.1198 & 0.119 & 0.1229 \\
82 & 2.3986 & 2.2900 & 2.28 & 2.2966 & 0.3719 & 0.3534 & 0.352 & 0.3556 \\
92 & 5.3212 & 4.9863 & 4.97 & 5.0035 & 0.8837 & 0.8214 & 0.820 & 0.8254 \\
100 & 9.8655 & 9.0688 & 9.05 & 9.0793 & 1.7471 & 1.5872 & 1.59 & 1.5904 \\
\midrule
& \multicolumn{4}{c}{2p$_{1/2}$} & \multicolumn{4}{c}{2p$_{3/2}$} \\
\cmidrule{2-5}\cmidrule{6-9}
$Z$ & {TW} & {Ref. \cite{Persson1993}} & {Ref. \cite{Mohr1998}} & {Ref. \cite{Ivanov2025}} & {TW} & {Ref. \cite{Persson1993}} & {Ref. \cite{Mohr1998}} & {Ref. \cite{Ivanov2025}} \\
\midrule
36 & 0.000070 & 0.000059 & 0.0001 &  & 0.000029 & 0.000020 & 0.0000 &  \\
54 & 0.0018 & 0.0016 & 0.0014 &  & 0.0005 & 0.0004 & 0.0004 &  \\
70 & 0.0164 & 0.0153 & 0.015 & 0.0163 & 0.0034 & 0.0028 & 0.003 & 0.0036 \\
82 & 0.0708 & 0.0661 & 0.067 & 0.0680 & 0.0115 & 0.0094 & 0.010 & 0.0109 \\
92 & 0.2193 & 0.2057 & 0.205 & 0.2091 & 0.0261 & 0.0226 & 0.021 & 0.0249 \\
100 & 0.5355 & 0.4978 & 0.496 & 0.5028 & 0.0498 & 0.0431 & 0.042 & 0.0465 \\
\bottomrule
\end{tabular*}
\end{table}

\begin{table}[!t]
\caption{Self energy correction to selected subshells for $Z=50$ and $Z=90$ hydrogenlike ions calculated with point-like nucleus model, in terms of $F(Z\alpha)$, compared to reference ones from Refs.~\cite{Mohr1992a} and \cite{Yerokhin2025} (see main text for details).\label{tab:bech_h-like}}
\tabcolsep=0.4\tabcolsep
\centering
\begin{tabular}{@{}@{\extracolsep{\fill}} l *6{S} @{}}
\toprule
& \multicolumn{3}{c}{$Z=50$} & \multicolumn{3}{c}{$Z=90$}\\
\cmidrule{2-4}\cmidrule{5-7}
Subshell & {Present} & {Reference} & {difference \%} & {Present} & {Reference} & {difference \%} \\
\midrule
1s & 1.86408 & 1.86427 & -0.01 & 1.48744 & 1.48754 & -0.01 \\
2s & 2.22421 & 2.22434 & -0.01 & 2.16653 & 2.16688 & -0.02 \\
3s & 2.27510 & 2.27572 & -0.03 & 2.19303 & 2.19356 & -0.02 \\
4s & 2.28421 & 2.28422 & 0.00 & 2.15772 & 2.15818 & -0.02 \\
5s & 2.28395 & 2.28333 & 0.03 & 2.12226 & 2.12251 & -0.01 \\
2p$_{1/2}$ & 0.00805 & 0.00801 & 0.45 & 0.29350 & 0.29307 & 0.15 \\
3p$_{1/2}$ & 0.04116 & 0.04136 & -0.48 & 0.36573 & 0.36601 & -0.08 \\
4p$_{1/2}$ & 0.05405 & 0.05377 & 0.53 & 0.38153 & 0.38190 & -0.10 \\
5p$_{1/2}$ & 0.06046 & 0.05966 & 1.35 & 0.38511 & 0.38487 & 0.06 \\
2p$_{3/2}$ & 0.19987 & 0.20005 & -0.09 & 0.29079 & 0.29067 & 0.04 \\
3p$_{3/2}$ & 0.22109 & 0.22145 & -0.16 & 0.33432 & 0.33497 & -0.19 \\
4p$_{3/2}$ & 0.23026 & 0.22993 & 0.14 & 0.35044 & 0.35071 & -0.08 \\
5p$_{3/2}$ & 0.23477 & 0.23413 & 0.27 & 0.35760 & 0.35746 & 0.04 \\
3d$_{3/2}$ & -0.03756 & -0.03777 & -0.56 & -0.02249 & -0.02248 & 0.03 \\
4d$_{3/2}$ & -0.03441 & -0.03480 & -1.11 & -0.01471 & -0.01488 & -1.17 \\
5d$_{3/2}$ & -0.03266 & -0.03312 & -1.41 & -0.01068 & -0.01080 & -1.19 \\
3d$_{5/2}$ & 0.04719 & 0.04755 & -0.76 & 0.06119 & 0.06117 & 0.03 \\
4d$_{5/2}$ & 0.05225 & 0.05073 & 2.99 & 0.06720 & 0.06734 & -0.20 \\
5d$_{5/2}$ & 0.05276 & 0.05241 & 0.67 & 0.07030 & 0.07037 & -0.09 \\
4f$_{5/2}$ & -0.02002 & -0.02066 & -3.09 & -0.01886 & -0.01901 & -0.78 \\
5f$_{5/2}$ & -0.02005 & -0.01993 & 0.59 & -0.01830 & -0.01792 & 2.11 \\
4f$_{7/2}$ & 0.02089 & 0.02161 & -3.33 & 0.02392 & 0.02434 & -1.73 \\
5f$_{7/2}$ & 0.02168 & 0.02245 & -3.39 & 0.02543 & 0.02570 & -1.07 \\
\bottomrule
\end{tabular}
\end{table}

\begin{table}[!t]
\caption{Comparison between experimental and theoretical energies (in eV) for the \cf{1s\ 2p\ ^{1}P_{1} \to 1s^2\ ^{1}S_{0}} transition for selected He-like ions. Present calculations, performed by the original \textsc{Grasp2018} \texttt{rci} program (labeled GO) and by a new \texttt{rci-q} program presented in this work (labeled TW), are presented. The numbers in parentheses represent the uncertainty of number quoted. The weighted mean of experimental results are also calculated. \label{tab:bech_he-like}}
\centering
\tabcolsep=0.5\tabcolsep
\begin{tabular*}{\linewidth}{@{}@{\extracolsep{\fill}} l *4{S[table-format=7.3]} S[table-format=7.6] r @{}}
\toprule
Z & {TW} & {GO} & {Artemyev} & {Plante} & {Exp.} & Ref. \\
& & & {et al. \cite{Artemyev2005}} & {et al. \cite{Plante1994}} & & \\
\midrule
26 & 6700.455(14) & 6700.441(14) & 6700.435(1) & 6700.423 & 6700.762(0.362) & \cite{Aglitskii1989} \\
 &  &  &  &  & 6700.725(0.201) & \cite{Beiersdorfer1989} \\
 &  &  &  &  & 6700.441(0.049) & \cite{Kubicek2014} \\
 &  &  &  &  & 6700.90(0.25) & \cite{Briand1984} \\
 &  &  &  &  & 6700.549(0.070) & \cite{Rudolph2013} \\
 &  &  &  &  & 6700.499(0.039) & mean \\
36 & 13114.474(13) & 13114.434(13) & 13114.471(4) & 13114.411 & 13115.45(0.30) & \cite{Indelicato1986a} \\
 &  &  &  &  & 13114.68(0.36) & \cite{Widmann1996} \\
 &  &  &  &  & 13114.47(0.14) & \cite{Epp2015} \\
 &  &  &  &  & 13113.80(1.20) & \cite{Briand1984a} \\
 &  &  &  &  & 13114.64(0.12) & mean \\
54 & 30630.256(15) & 30630.014(15) & 30630.051(17) & 30629.667 & 30629.1(3.5) & \cite{Briand1989} \\
 &  &  &  &  & 30631.2(1.2) & \cite{Thorn2009} \\
 &  &  &  &  & 30631.0(1.1) & mean \\
92 & 100619.925(16) & 100609.200(16) & 100610.89(65) & 100613.924 & 100626(35) & \cite{Briand1990} \\
 &  &  &  &  & 100598(107) & \cite{Lupton1994} \\
 &  &  &  &  & 100623(33) & mean \\
\bottomrule
\end{tabular*}
\end{table}

\section{Benchmarking}
\label{sec:benchmark}

\subsection{H-like systems}

In order to check the fitting procedure quality, the calculated self energy correction to particular subshells of hydrogenlike ions were compared to reference values -- see Table~\ref{tab:bech_h-like}. One can see that for the s shells the difference is up to 0.03\%. For (2--4)p shells the difference is up to 0.5\% and for the other shells the difference is up to 3\% for $Z=50$ and up to 2\% for $Z=90$.

\subsection{He-like systems}

Heliumlike atomic systems are used for studying the QED effects for many years \cite{Indelicato2019}. In Table~\ref{tab:bech_he-like} the present calculated energies of for the \cf{1s\ 2p\ ^{1}P_{1} \to 1s^2\ ^{1}S_{0}} (the `w' or $K\alpha_1$ line) transition for selected He-like ions are presented and compared to experimental and other theoretical numbers. In order to account the electron correlation effects, the \textsc{Grasp2018} calculations were performed within configuration-interaction mode with active space obtained by single and double excitations from the reference \cf{1s\ 2p} and \cf{1s^2} configurations to the virtual orbitals up to $n=6$ and $l=3$. The $\omega>0$ part of the Breit interaction has been also included. The uncertainties of calculated results are estimated as $\Delta E = \left[ (E(n\le6,l\le3)-E(n\le5,l\le3))^2+(E(n\le6,l\le3)-E(n\le6,l\le2))^2 \right]^{1/2}$, counting effects of extending the active space by increasing both the $n$ and $l$ numbers.. 
One can see the difference between original \textsc{Grasp2018} QED model numbers (labeled GO) and the present model ones (labeled TW) is small for lower $Z$ atoms but rises to 10~eV for $Z=92$. In all cases the present QED model reproduce weighted mean of experimental numbers better than original \textsc{Grasp2018} QED model. For $Z=92$ the TW number is even closer to center of uncertainty interval of the experimental result than the Artemyev et al. \cite{Artemyev2005} and Plante et al. \cite{Plante1994} results. Unfortunately, the current precision of measurements of $K\alpha$ lines for heavy atoms is not enough to examine QED models for sure.

\subsection{F-like systems}

The \cf{2p_{3/2}-2p_{1/2}} fine splitting within the F-like isoelectronic sequence (\cf{1s^2\ 2s^2\ 2p^5}) has been recently suggested to be a highly accurate test of Breit and QED effects \cite{Li2018b,Volotka2019,Shabaev2020}, due to fact that for high-$Z$ elements the correlation contribution to the \cf{2p_{3/2}-2p_{1/2}} fine splitting energy is much smaller compared to the Breit and QED corrections \cite{Li2018b}. The fine splittings for uranium and molybdenum F-like ions have been chosen to test QED model in the \textsc{rci-q} program, because for these ions the experimental data are measured with $<10\%$ accuracy, that gives an opportunity to examine QED models. The data are stored in Table~\ref{tab:bech_f-like} and compared to other theoretical results. The experimental QED contributions to the \cf{2p_{3/2}-2p_{1/2}} fine splitting were estimated by Li et al. \cite{Li2018b}, by subtracting the non-QED \textsc{Grasp2k} calculations from experimental measurements. 
One can see from Table~\ref{tab:bech_f-like} that for $Z=92$ the calculated QED contributions of Volotka et al. \cite{Volotka2019}, Shabaev et al. \cite{Shabaev2020}, and from present work are all close to the experimental value. However, both numbers provided by Li et al. \cite{Li2018b} are far away from uncertainty interval of experimental result.

\begin{table}[!t]
\caption{QED corrections to the \cf{2p_{3/2}-2p_{1/2}} fine splitting energy for $Z=42$ and $Z=92$ F-like ions in the units of eV.\label{tab:bech_f-like}}
\centering
\begin{tabular*}{\linewidth}{@{}@{\extracolsep{\fill}} l *2{S} @{}}
\toprule
& {$Z=42$} & {$Z=92$}\\
\midrule
Experiment$^\ast$, Li et al. \cite{Li2018b} & 0.220(20) & 2.25(16) \\
\multicolumn{3}{c}{Perturbatively calculated:}\\
\textsc{Grasp2k} original model, Li et al. \cite{Li2018b} & 0.203 & 1.33 \\
\textsc{Grasp2k} + \textsc{Qedmod}, Li et al. \cite{Li2018b} & 0.229 & 1.79 \\
Volotka et al. \cite{Volotka2019} & 0.222 & 2.47 \\
Shabaev et al. \cite{Shabaev2020} & 0.241(1) & 2.60(4) \\
This work & 0.232 & 2.47\\
\multicolumn{3}{c}{Self-consistently calculated:}\\
Shabaev et al. \cite{Shabaev2020} & 0.238 & 2.12 \\
\bottomrule
\end{tabular*}
\small 
$^\ast$ Deduced from experimental fine-structure energies and non-QED calculations performed by the \textsc{Grasp2k} code by Li et al. \cite{Li2018b}.
\end{table}

\section{Installation of \textsc{Rci-Q} program and merging with \textsc{Grasp2018}}

The \texttt{rci-q} program supersedes the \texttt{rci} program provided by the \textsc{Grasp2018} package. The following files are added/changed:
\begin{itemize}
\item \texttt{qed\_slfen\_pot.f90} is added as a replacement for \texttt{qed\_slfen.f90}, 
\item \texttt{vacpol.f90} is changed (\texttt{vac4} subroutine is commented and \texttt{vacpol\_wk} subroutine is used instead), 
\item \texttt{se\_pot\_el.f90}, \texttt{se\_pot\_el\_I.f90}, \texttt{se\_pot\_int.f90}, \texttt{se\_pot\_int\_I.f90}, \texttt{se\_pot\_low.f90}, \newline \texttt{se\_pot\_low\_I.f90}, \texttt{se\_pot\_mag.f90}, \texttt{se\_pot\_mag\_I.f90}, \texttt{se\_fs.f90}, \texttt{se\_fs\_I.f90}, \newline \texttt{vacpol\_wk.f90}, \texttt{vacpol\_wk\_I.f90} are added.
\end{itemize}

Assuming that the \textsc{Grasp2018} package has already been installed in the \texttt{GRASPDIR} directory, the installation procedure of the \textsc{Rci-Q} package is as follows. 
\begin{enumerate}
\item Download the \texttt{srcq} archive from the repository \url{https://gitlab.com/Koziol/rci-q}.
\item Unpack the \texttt{srcq} archive wherever, let us say to the \texttt{TMPDIR} directory.
\item Copy the \texttt{TMPDIR/rci-q} and \texttt{TMPDIR/rci-q\_mpi} directories to the \texttt{GRASPDIR/src/appl/} directory. For the installation convenience these directories contain also all files from the original \textsc{Grasp2018} package necessary to build \texttt{rci-q} and \texttt{rci-q\_mpi} executables. 
\item Under the \texttt{bash} shell perform the following command sequence, starting from the \texttt{GRASPDIR} working directory.
\begin{itemize}
\item \texttt{source make-environment\_xxx} (where \texttt{xxx} is the compiler name, for example \texttt{gfortran} -- see the \textsc{Grasp2018} package manual \cite{Jonsson2023} for details)
\item \texttt{cd src/appl/rci-q}
\item \texttt{make}
\item \texttt{cd ../rci-q\_mpi}
\item \texttt{make}
\end{itemize}
\end{enumerate}

The generated executable program files, \texttt{rci-q} and \texttt{rci-q\_mpi}, will be placed in the \texttt{GRASPDIR/bin} directory.


The \texttt{rci-q} and \texttt{rci-q\_mpi} programs do not require additional inputs. Just replace \texttt{rci} by \texttt{rci-q} (or \texttt{rci\_mpi} by \texttt{rci-q\_mpi}) in your scripts. 

The computational overhead, related to evaluating the Flambaum--Ginges potential on a radial grid, is about 20\%. A test case was a multi-configuration Dirac--Hartree--Fock and configuration interaction (MCDHF-CI) calculations involving \num{2.4e5} configuration state functions (CSFs) related to the ground state of Th$^{39+}$ ion \cite{Koziol2025a}. Calculation times using 16 Intel\textsuperscript{\tiny\textregistered} Xeon\textsuperscript{\tiny\textregistered} 2.60~GHz CPUs was: 52 minutes by using the \texttt{rci\_mpi} program and 63 minutes by using the \texttt{rci-q\_mpi} program. 
However, for practical reasons the overhead may be much smaller. Involving QED corrections within MCDHF-CI calculations for higher virtual orbitals might be problematic, as these virtual orbitals might be far from the real counterparts. So, a common approach is to apply QED corrections in calculations performed for a small CSFs base and then shift the numbers obtained in much larger CSFs base.

\section{Conclusions}

The \textsc{Rci-Q} package, being an extension to the \textsc{Grasp2018} suite, has been presented. Benchmark tests show improvement in the high-$Z$ regime, where QED effects are significant, while for lower-$Z$ atoms the difference between results obtained by the \texttt{rci-q} program and original \textsc{Grasp2018} \texttt{rci} program is much smaller.

\section*{Acknowledgments}
The author gratefully acknowledges (mentioned in alphabetical order) 
Gustavo A. Aucar, I. Agust\'{\i}n Aucar, Robert Berger, Konstantin Gaul, and Kjell Janke 
for helpful discussions.


%

\end{document}